Excess energy fluctuations with applications to deviations from Debye's specific heat

Ralph V. Chamberlin[1,*] and Sumiyoshi Abe[2,3,4]

[1]Department of Physics, Arizona State University, Tempe, AZ 85287-1504, USA

[2]Department of Physics, College of Information Science and Engineering, Huaqiao University, Xiamen 361021, China

[3]Institute of Physics, Kazan Federal University, Kazan 420008, Russia

[4]Department of Natural and Mathematical Sciences, Turin Polytechnic University in Tashkent, Tashkent 100095, Uzbekistan

**Abstract** Measured specific heats often exceed Debye's $T^3$-law, even in high-purity single crystals. Analogous excess energy fluctuations are found in molecular dynamics (MD) simulations of crystals with no defects. Here, a theory is developed for the fluctuations based on rapid modulation of the ground-state energy due to the motion of next-nearest-neighbor atoms. Crucially, the modulations must be adiabatic, with time- and phase-averaging before thermal averaging, consistent with evidence from the simulations and multiple experimental techniques showing that localized excitations are decoupled from the heat bath. Emergent nonextensivity is interpreted by analogy with anomalous diffusion. The theory modifies the standard relation between energy fluctuations and specific heat, giving good agreement with the simulations and new insight into many measurements. The theory may also provide a basis for understanding excess specific heat in amorphous materials and anomalous noise in quantum devices.

*Corresponding author at: Department of Physics, Arizona State University, Tempe AZ 85287-1504, USA. *E-mail address:* ralph.chamberlin@asu.edu (R. V. Chamberlin).

## 1. Introduction

The standard theory for describing specific heat, $C$, in crystals at low temperatures ($T \ll 300$ K) is Debye's treatment of quantized harmonic waves (phonons) [1,2]. Debye's theory gives $C \propto T^3$ as $T \to 0$. The accuracy of this theory should increase with decreasing $T$, but many measurements, even on ultra-pure single-crystal dielectrics, show excess $C$ at $T \ll 1$ K [3-5]. The authors of Ref. [3] attribute this excess to their technique of measuring time-resolved specific heat as being "…a very sensitive test for residual impurities in perfect crystals." However, molecular dynamics (MD) simulations of crystals with *no impurities* show an analogous excess in energy fluctuations [6]. Here, we develop a theory that quantitatively explains the excess energy fluctuations in the simulations and gives novel insight into measured behavior.

Canonical-ensemble theory contains a formal identity connecting specific heat and energy fluctuations [7], which we call the $C \leftrightarrow (\Delta E)^2$ relation. The theory we propose adds a correction to this relation. The correction involves the linear term in the potential energy between next-nearest-neighbor (nnn) atoms that rapidly modulates the quadratic term between nearest-neighbor (nn) atoms. The net behavior comes from time- and phase-averaging of the fast modulations, then thermal averaging of the quadratic interactions. The key point is that uncorrelated motions of nnn atoms are fast enough to effectively conserve *local energy* within a small surrounding region, essentially decoupled from the large heat bath, consistent with evidence from the simulations [6] and many measurements [3-5,8-10]. We first treat the modulations classically, as needed for MD simulations. Then, we extend our theory to the quantum regime using a semiclassical treatment of the modulations. We find that such an approach modifies the density of states in Debye's theory in a way that is consistent with the excess specific heat measured in many materials.

Excess specific heat is ubiquitous in glasses [1,11]. This excess has been attributed to defects that yield tunneling two-level systems (TTLS) [12-16]. Similarly, excess noise in quantum devices, including qubits and single-particle detectors [17-21], has also been attributed to TTLS. However, the source of TTLS in most materials remains unknown [14,15], with behavior that may already be refuted by experiment, as emphasized in Ref. [22]. Our theory and semiclassical picture provide qualitative insight into the excess specific heat measured in glasses, where interaction potentials between most atoms have nonzero slopes, so that the motion of every atom strongly modulates the energies of most neighbors.

The remainder of this paper is organized as follows. Section 2 presents the experiments and simulations that guide our theory. Section 3 develops the theory for how fluctuating atoms modulate the interaction energies in classical MD simulations. Section 4 extends this classical theory to a semiclassical treatment that is applied to Debye's theory of specific heat. Section 5 finishes with some conclusions.

**2. Background: Experiments and simulations**

The main experiment that guides our theory is time-resolved specific heat [3-5]. When a short heat pulse is applied to a high-purity single crystal at $T \ll 1$ K, the short-time (initial) specific heat agrees with Debye's theory, but the long-time (equilibrium) value can be much higher. This excess specific heat is qualitatively similar to the behavior of glasses. Indeed, additional insight into our theory comes from experiments showing that the excess specific heat in glasses involves dynamic heterogeneity due to localized normal modes [23- 28], with analogous heterogeneity measured in crystals [29-31]. The anharmonic interactions that cause these heterogeneities yield the boson peak indicative of excess high-frequency vibrations in glasses, and in crystals [32]. Further evidence comes from neutron scattering measurements that show a sharp reduction in the correlations between nnn atoms compared to nn atoms [33], as needed for the relatively independent motion of nnn atoms in our theory.

In Fig. 1, we present results from measurements [3-5] of time-resolved specific heat on five high-purity single crystals. The inset of Fig. 1 (A) shows a sketch of the "box" model for interpreting the behavior [10,31]. An electrical pulse containing a fixed amount of heat is applied to the resistor on one side of the sample, with temperature change ($\delta T$) measured as a function of time on the other side. The round "boxes" in the sketch represent sets of localized degrees of freedom that have local temperatures ($T_i$) and heat resistors (time constant, $\tau_i$) coupling them weakly to the heat bath. The measurements (open circles) in Fig. 1 (A) are on a quartz crystal at a base temperature of $T \sim 0.2$ K. Note how the change in temperature, $\delta T$, starts at zero, rises sharply until it reaches a peak at $t \sim 0.01$ ms, then recovers towards zero in multiple stages. The overshoot and peak at ~0.01 ms are due to the initial arrival of ballistic phonons from the heater. After the peak, the added heat becomes evenly dispersed among all phonons that form the heat bath, allowing $\delta T$ to stabilize at an initial plateau (dotted line) that gives $\delta T_D$ from Debye's theory. However, starting at $t \sim 1$ ms, $\delta T$ relaxes again, eventually reaching an equilibrium plateau (dashed line) at $t > 10$ ms which gives the total specific heat of the sample. Finally, at much longer times ($t \gg 100$ ms, not shown), $\delta T$ returns to zero as the excess heat flows out of the sample through the wires that connect it to the cryostat.

Because the specific heat is inversely related to $\delta T$ [5], Fig. 1 (A) shows that the equilibrium specific heat is significantly higher than its initial value. The initial value of $C$ comes from the homogeneous heat bath of phonons that yield Debye's theory, while the equilibrium value has additional degrees of freedom. In fact, during the nearly-exponential relaxation connecting these plateaus (solid curve between $0.1 \lesssim t \lesssim 10$ ms), the excess heat supplied to the sample is not flowing across the sample (heat bath equilibration occurs at $t \approx 0.01$ ms), and it is not flowing out of the sample (coupling to the cryostat occurs at $t \gg 100$ ms). Thus, the excess heat must be flowing from the heat bath into other degrees of freedom inside the sample. Net heat flows only if there is a temperature difference. Hence, this experiment establishes that there are multiple temperatures inside a sample that do not equilibrate to a common $T$ until $t > 10$ ms. Therefore, while these localized degrees of freedom are relaxing, their thermal exchange with the heat bath is too slow to be governed by a single $T$. In other words, fast fluctuations of these localized degrees of freedom are *decoupled from the heat bath* due to their relatively weak thermal link, represented by heat resistors in the sketch in Fig. 1 (A). Figure 1 (B) shows the $T$ dependence of the equilibrium specific heat, from time-resolved measurements on five high-purity single crystals given in the legend [3-5]. The measurements, which are normalized by Debye's specific heat ($C_D T^3$), show clear deviations from Debye's $T^3$-law at low $T$.

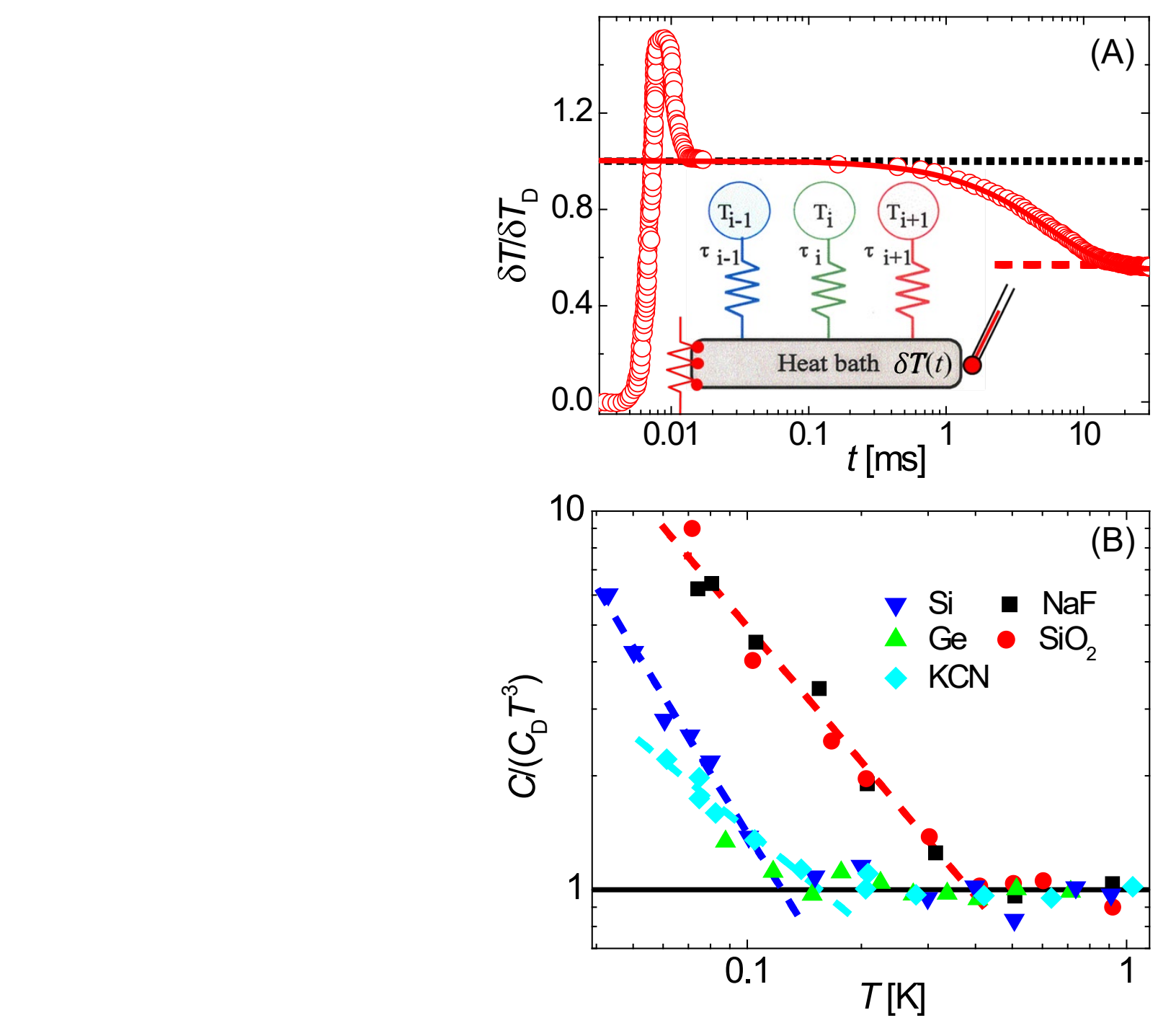

**Fig. 1.** Measurements yielding specific heat as a function of (A) time and (B) temperature [3-5]. (A) Symbols show normalized temperature change, $\delta T$ divided by Debye's $T^3$-law, $\delta T_D$, from measurements of an $SiO_2$ crystal at $T{\sim}0.2$ K. Lines show $\delta T_D$ (dotted), the equilibrium value (dashed), and an exponential fit to the relaxation (solid). {Inset: sketch showing how the measurement utilizes a heat pulse applied to one side of the sample, while $\delta T$ is measured on the other side, with relaxation on the 1 to 10 ms time scale attributed to heat flowing through the heat resistors (time constants $\tau_i$) into slow degrees of freedom (boxes with temperatures $T_i$) [10]}. (B) Symbols show the equilibrium specific heat as a function of temperature, normalized by Debye's specific heat [solid line at $C/(C_D T^3) = 1$] from five high-purity single crystals, given in the legend. Dashed lines as $T \to 0$ show power-law fits to the excess specific heats, $C/(C_D T^3) - 1 = \alpha/T^\mu$. Empirical exponents range from $\mu = 0.94 \pm 0.15$ for KCN (cyan) to $1.64 \pm 0.12$ for Si (blue), with an average of $\mu = 1.4 \pm 0.3$.

Figure 2 (A) is a sketch showing how a large simulation can be subdivided into smaller subsystems (blocks) to study local thermal behavior. Specifically, the fixed volume and number of particles in each block should yield canonical-ensemble behavior, with the many distant blocks providing the large heat bath needed to define $T$. However, MD simulations of several models [6] exhibit significant deviations from the predictions of the canonical ensemble, especially at low $T$.

The main part of Fig. 2 (B) shows results from MD simulations of the Lennard-Jones (L-J) model as a function of temperature for three different block sizes, given in the legend, with parameters chosen for argon [34]. The notation used is as follows. Let $q$ be potential ($u$) or kinetic ($k$) energy per atom in a block, so that $q$ is nominally intensive. Averages of $q$, from time averaging combined with the arithmetic mean over all blocks, are denoted by $\langle q\rangle_{t,bl}$. Energy differences with temperature (subscript $T$) are given by $\delta_T q \equiv q(T+\delta T) - q(T)$, while energy variances due to fluctuations in time are $\left(\Delta_{t,bl} q\right)^2 \equiv \langle\left(q(t) - \langle q\rangle_{t,bl}\right)^2\rangle_{t,bl}$. The law of equipartition of energy predicts that each quadratic degree of freedom will have a $T$-dependent component equal to $k_B T/2$. Indeed, in Fig. 2 (B) from $T < 10$ K, the black line near unity gives $\delta_T\langle u\rangle_{t,bl}/\delta_T\langle k\rangle_{t,bl} = 0.9998 \pm 0.0013$, which, from the virial theorem, establishes that the net interactions are accurately characterized by a quadratic potential. However, for fluctuations, the open symbols in Fig. 2 (B) show that the ratio, $\left(\Delta_{t,bl} u\right)^2/$ $\left(\Delta_{t,bl}k\right)^2$, increases rapidly as $T \rightarrow 0$, reaching 10-100 times larger than unity predicted by the standard $C \leftrightarrow (\Delta E)^2$ relation. For blocks of $n$ atoms, these excess fluctuations are empirically characterized by [6]

$$\frac{\left(\Delta_{t,bl}u\right)^2}{\left(\Delta_{t,bl}k\right)^2} - 1 = \eta\, n^{\gamma-1}\epsilon_{LJ}/(k_B T). \quad (1)$$

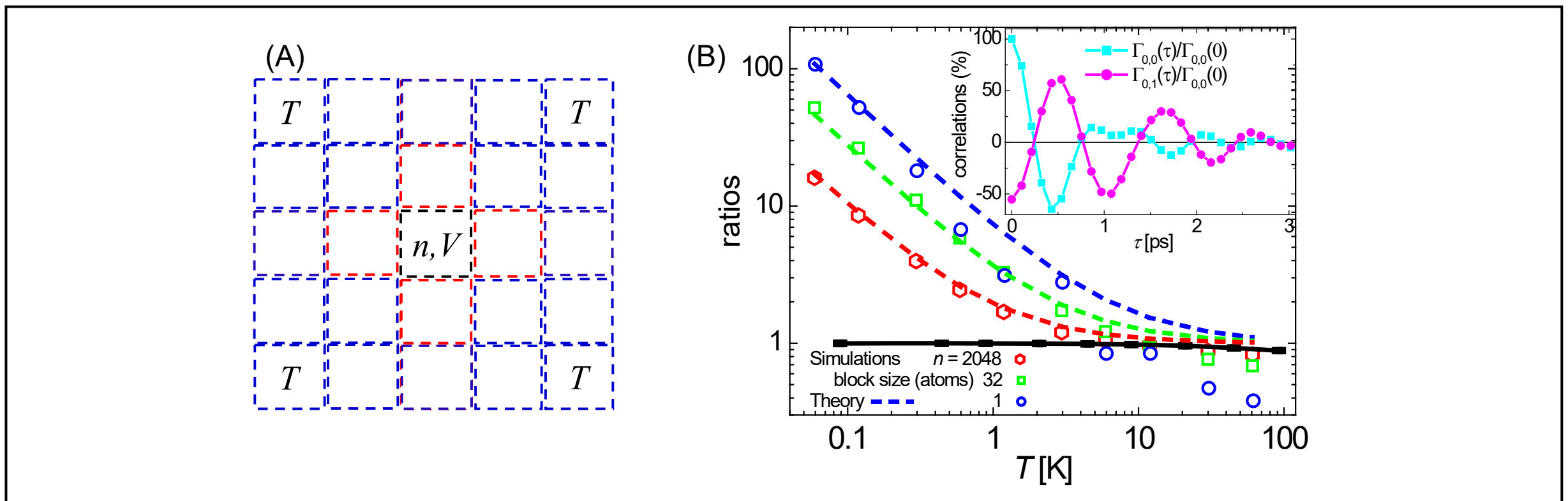

**Fig. 2.** (A) Sketch of a system subdivided into small blocks. In 3-dimensions, each block is cube-shaped, with fixed volume ($V$) and number of atoms ($n$). The many distant blocks serve as a heat bath, defining temperature ($T$) for the canonical ensemble. (B) Results from MD simulations of 442,368 Lennard-Jones (L-J) atoms using parameters for argon [6]. Main figure shows ratios from potential energy ($u$) divided by kinetic energy ($k$) as a function of $T$. The black line (with error bars) gives the ratio of average energies, $\delta_T\langle u\rangle_{t,bl}/\delta_T\langle k\rangle_{t,bl} \equiv \left[\langle u\rangle_{t,bl}(T+\delta T) - \langle u\rangle_{t,bl}(T)\right]/\left[\langle k\rangle_{t,bl}(T+\delta T) - \langle k\rangle_{t,bl}(T)\right]$. (See text for details about notation.) This ratio's proximity to unity ($0.9998 \pm 0.0013$) establishes the accuracy of the harmonic approximation for potential energy. Open symbols show the ratio of fluctuations in energy, $\left(\Delta_{t,bl} u\right)^2/\left(\Delta_{t,bl} k\right)^2$. As $T \to 0$, these fluctuations diverge from the canonical-ensemble prediction of unity, demonstrating the failure of the $C \leftrightarrow (\Delta E)^2$ relation. The blue dashed line shows the proposed theory that uses time- and phase-averaging followed by thermal averaging to yield $\eta = 0.053833$ in Eq. (1). Then, adding anomalous diffusion for $\gamma = 3/4$, Eq. (1) yields the dashed lines that match the energy fluctuations of blocks having $n$=32 (green) and 2048 (red). [Inset: autocorrelations (cyan squares, normalized to 100% at $t = 0$) and pair-correlations (magenta circles) between all 6 adjacent blocks (4 red blocks in (A)), each containing $n = 4$ atoms.]

Here, $\epsilon_{LJ}/k_B = 118.2$ K gives the characteristic energy of the L-J model for argon [34], with $\gamma = 0.75 \pm 0.05$ and $\eta = 0.054 \pm 0.004$ found from best fits to the simulations [6].

The inset of Fig. 2 (B) clarifies the primary cause of the excess fluctuations. We denote the autocorrelation function within all blocks by $\Gamma_{0,0}(\tau) \equiv \langle u_0(t)u_0(t+\tau)\rangle_{t,bl}$, and the pair-correlation function between each block and its six adjacent blocks [four such blocks are shown in red in Fig. 2 (A)] by $\Gamma_{0,1}(\tau) \equiv \langle u_0(t)u_1(t+\tau)\rangle_{t,bl}$. Again, the angle bracket and subscript denote the time average for each block (or for each set of neighboring blocks) followed by the arithmetic mean over all blocks. The ratio, $\Gamma_{0,0}(\tau)/\Gamma_{0,0}(0)$, starts at 100% and exhibits roughly damped harmonic relaxation with a period of $\approx 1$ ps. While $\Gamma_{0,1}(\tau)/\Gamma_{0,0}(0)$ starts at $\lesssim -50\%$ and is primarily anticorrelated with $\Gamma_{0,0}(\tau)/\Gamma_{0,0}(0)$. Thus, more than 50% of the excess energy in each central block during its initial fluctuation comes from its six adjacent blocks. Hence, it can be assumed that essentially 100% of the excess energy in the initial fluctuation comes from the surrounding "shell" of $3^3 - 1 = 26$ blocks that touch each central block, not from the large number of distant blocks needed for a well-defined $T$. In fact, during the first oscillation at $t \approx 0.5$ ps, the auto- and pair-correlations are equal in magnitude but opposite in sign, showing that most of the excess energy in the initial fluctuation remains in the six neighboring blocks. Indeed, local energy is conserved by trading back-and-forth between each block and its surrounding shell. These oscillations are too fast for their energy to reach beyond the "local bath" of surrounding blocks; the fluctuating energy is effectively decoupled from the large heat bath needed for the canonical ensemble.

The block-size dependence of energy fluctuations in Fig. 2 (B) is clearly nonextensive. Although surface effects yield nonextensive energies when atoms are near the surface of an isolated small system, four facts establish that surface effects cannot explain the behavior shown in Fig. 2 (B). First, there are *no* surface atoms; each block is a virtual construct inside a bulk simulation, so that every atom interacts with the same number of neighboring atoms. Second, the simulations yield Eq. (1) with $n^{0.75}$, not $n^{2/3}$ needed for surface effects. Third, the $C \leftrightarrow (\Delta E)^2$ relation does not depend on the nature of the interaction: if average energies are described by the canonical ensemble, any surface effect in the fluctuations must also occur in the specific heat, whereas Fig. 2 (B) shows equal potential and kinetic parts of the specific heat, but unequal fluctuations. Finally, as shown in Fig. 3 (B), although the interaction energy is dominated by nn atoms, excess energy fluctuations occur only if there is a nonzero slope in the potential term between nnn atoms. Therefore, the nonextensive size dependence cannot come from surface effects.

Key insight into the cause of the breakdown in the $C \leftrightarrow (\Delta E)^2$ relation comes from Fig. 3, where the behavior of the L-J model is shown as a function of interaction cutoff radius, $r_c$. Lines in Fig. 3 (A) show excellent agreement between specific heat and the law of equipartition of energy. Indeed, from kinetic energy (black), $\delta_{\mathrm{T}}\langle k\rangle_{t,bl}/\delta(k_B\mathrm{T}) = 1.4992 \pm 0.0007$ is well-within statistical uncertainty of 3/2 expected in 3 dimensions. Similarly, from potential energy (red), $\delta_{\mathrm{T}}\langle u\rangle_{t,bl}/\delta(k_B\mathrm{T}) = 1.494 \pm 0.002$, but only if $r_c/\sigma \geq 1.3$ needed for a well-defined quadratic potential between nn atoms, see red circle and dashed curve in Fig. 3 (D).

Symbols in Fig. 3 (B) show dimensionless energy fluctuations. From the kinetic energy (black diamonds) $n\left(\Delta_{t,bl}k\right)^2/(k_BT)^2 = 1.43 \pm 0.03$ (the factor $n$ is needed to make the per-atom variance intensive), in approximate agreement with the $C \leftrightarrow (\Delta E)^2$ relation. However, there are clear deviations from the $C \leftrightarrow (\Delta E)^2$ relation for potential energy. Specifically, when $1.3 \leq r_c \leq 1.6$, $n\left(\Delta_{t,bl}u\right)^2/(k_BT)^2 = 0.929 \pm 0.012$ is significantly below the value of 3/2 expected from the $C \leftrightarrow (\Delta E)^2$ relation. Most striking, however, is the divergent behavior when $r_c > 1.6$. Figure 3 (D) shows the reason. When $1.3 \leq r_c \leq 1.6$, the cutoff limits the interaction to a robust quadratic potential (red dashed curve) between nn atoms (red circle), whereas when $r_c > 1.6$, there is also a tangentially linear potential (blue dashed line) between nnn atoms (blue circle). Although the onset of a nonzero slope in the potential between nnn atoms has no perceptible influence on the specific heat in Fig. 3 (A), it conspicuously alters the energy fluctuations in Fig. 3 (B). Another signature of this linear potential is shown by the black squares in Fig. 3 (C), where the correlation between potential energy in neighboring blocks changes abruptly from positive for $r_c \leq 1.6$ to negative for $r_c > 1.6$. The inset of Fig. 2 (B) shows that this initial negative correlation persists as a function of time.

To appreciate the significance of deviations from the $C \leftrightarrow (\Delta E)^2$ relation it is useful to review its derivation. The notation we use, similar to that given in the paragraph preceding Eq. (1), is as follows. Let $Q$ represent the potential energy $U$, or kinetic energy $K$, in a block of $n$ atoms. Thus, $Q = nq$ is nominally extensive in $n$, unlike the per-atom quantities ($q = u, k$) used in Figs. 2 and 3. Again, the angled brackets with subscripts, $\langle Q\rangle_{t,bl}$, denote the time average of each block in the simulation combined with the arithmetic mean over all blocks, while $\delta_T$ denotes the temperature difference $\delta_T Q \equiv Q(T+\delta T) - Q(T)$. The time-averaged variance is $\left(\Delta_{t,bl}Q\right)^2 \equiv \langle\left(Q(t) - \langle Q\rangle_{t,bl}\right)^2\rangle_{t,bl}$. Now, let angled brackets with subscript $th$, $\langle Q\rangle_{th}$, denote the canonical-ensemble thermal average over all states in any block, so that the thermal-averaged variance is $(\Delta_{th}Q)^2 \equiv \langle (Q - \langle Q\rangle_{th})^2\rangle_{th}$. From ergodicity, the time- and thermal-averaged variances should equal each other, but Figs. 2 and 3 show clear deviations. We seek to understand why.

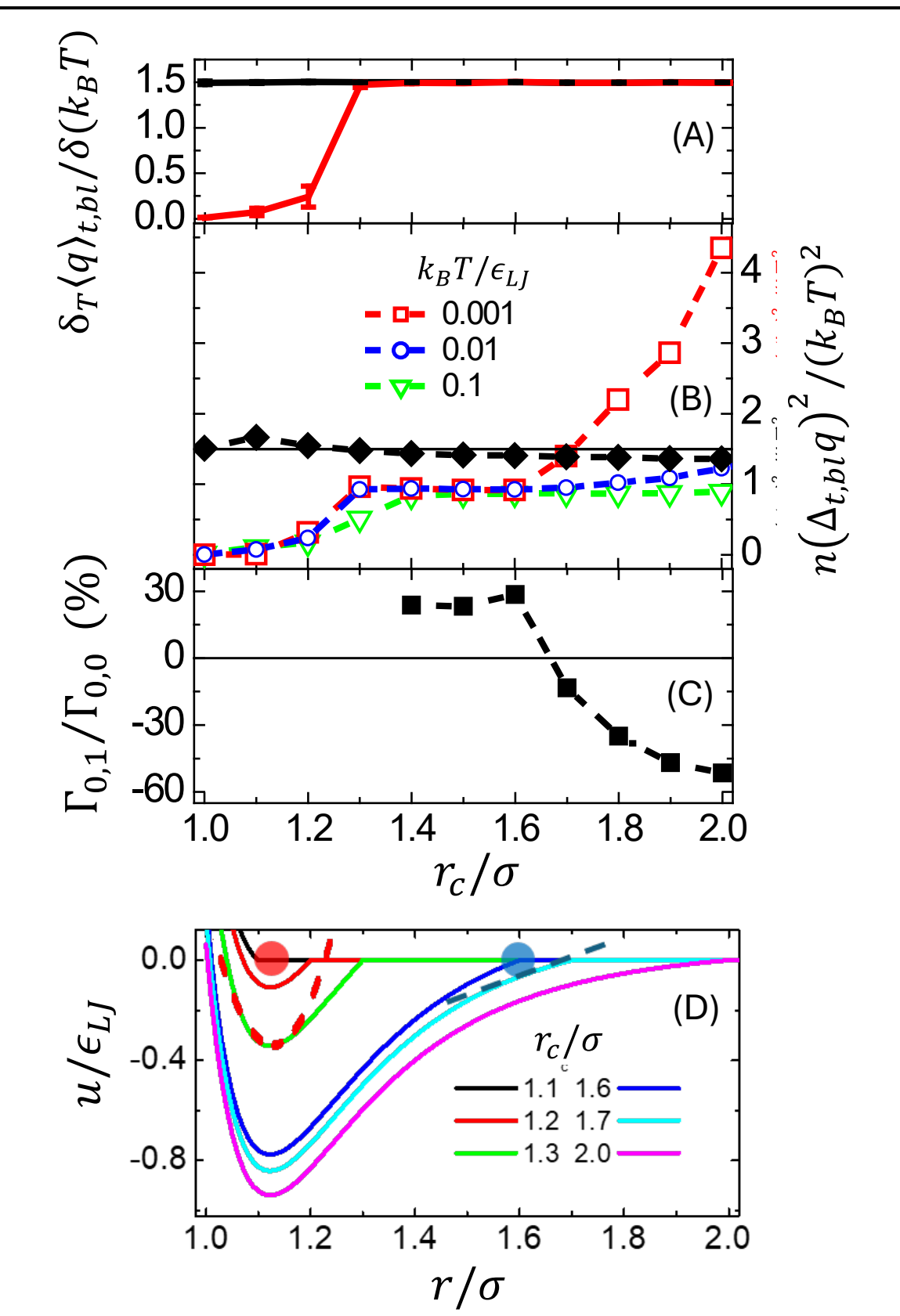

**Fig. 3.** Interaction cutoff radius ($r_c$) dependence of the L-J model from MD simulations [6]. (A) Lines show specific heats (with error bars visible when larger than the line width) from $q = k$ (black) and $q = u$ (red). Kinetic energies, averaged over 6 temperatures ($0.00005 \leq kT/\epsilon_{LJ} \leq 0.02$) and 13 cutoff radii ($1.0 \leq r_c/\sigma \leq 2.5$), give excellent agreement with the equipartition of energy: $\delta_T\langle k\rangle_{t,bl}/\delta(k_BT) = 1.4992 \pm 0.0007$. Potential energies, averaged over the same temperatures but limited to cutoffs having quadratic interactions between nn atoms [$r_c/\sigma > 1.2$, see (D)], give $\delta_T\langle u\rangle_{t,bl}/\delta(k_BT) = 1.494 \pm 0.002$. (B) Symbols show dimensionless fluctuations in kinetic energy (filled) and potential energy (open) for temperatures given in the legend. The kinetic energies yield $(\Delta_{t.bl}k)^2/(k_BT)^2 = 1.43 \pm 0.03$, consistent with the $C \leftrightarrow (\Delta E)^2$ relation. However, the potential energies deviate from this $C \leftrightarrow (\Delta E)^2$ relation, especially when the interaction includes a nonzero slope in the potential between nnn atoms [$r_c/\sigma > 1.6$, see (D)]. (C) Symbols show the normalized pair correlation in potential energy between neighboring blocks $\Gamma_{0,1}(0)/\Gamma_{0,0}(0)$. (D) Lines show potential energy for six cutoff radii, given in legend. Potential for $r_c/\sigma = 1.3$ (green) is the shortest cutoff radius that gives a well-defined quadratic potential (red dashed curve) between nn atoms (red circle). Potential for $r_c/\sigma = 1.7$ (cyan) is the shortest cutoff radius having a non-zero tangential slope (blue dashed line) between nnn atoms (blue circle).

From ergodicity for the averages of $Q$, we have

$$\langle Q\rangle_{t,bl} = \langle Q\rangle_{th}. \tag{2}$$

To repeat: the left-hand side is the time average (combined with block average) from simulations, while the right-hand side is the thermal average in the canonical ensemble. The validity of Eq. (2) is verified for both $U$ and $K$ by their agreement with 3/2 for the per-atom thermal averages in Fig. 3 (A). Thus, Eq. (2) is confirmed for averages, but what about fluctuations? Taking the temperature derivative of Eq. (2) and using the canonical-ensemble identity, $\delta_T\langle Q\rangle_{th}/\delta(k_BT) = (\Delta_{th}Q)^2/(k_BT)^2$, we have

$$\frac{\delta_T\langle Q\rangle_{t,bl}}{\delta(k_BT)} = \frac{(\Delta_{th}Q)^2}{(k_BT)^2}. \tag{3}$$

The crucial question is whether $(\Delta_{th}Q)^2 = \left(\Delta_{t,bl}Q\right)^2$ holds? That is,

$$\frac{\delta_T\langle Q\rangle_{t,bl}}{\delta(k_BT)} \stackrel{?}{=} \frac{\left(\Delta_{t,bl}Q\right)^2}{(k_BT)^2}. \tag{4}$$

At least for MD simulations of several models, Ref. [6] establishes that the answer is no. Indeed, for the L-J model, solid lines in Figs. 2 (B) and 3 (A) show excellent agreement with Eq. (2) for both $Q = K$ and $Q = U$, whereas the symbols in Figs. 2 (B) and 3 (B) clearly deviate quantitatively and qualitatively from Eq. (4) for $Q = U$. Thus, time-average fluctuations in MD simulations do not match thermal-average fluctuations. In other words, having equivalence between time- and thermal-averages does not assure equivalence in their fluctuations.

**3. Classical theory: Time-modulated quadratic potentials**

For many MD simulations [6], including those in Figs. 2 and 3, the excess energy fluctuations have three key features that infer their underlying mechanism. First, as shown in the inset of Fig. 2 (B) and black squares in Fig. 3 (C), essentially all the energy in the excess fluctuation comes from the local shell of neighboring blocks, not from the large heat bath needed to define $T$. Thus, these excess fluctuations involve motions that are essentially decoupled from the heat bath. Therefore, the fluctuations are controlled by conservation of local energy, not by a well-defined $T$. Second, the excess fluctuations require interactions beyond nn atoms, arising abruptly (but smoothly) when the interaction starts to include nnn atoms, as shown in Fig. 3 (B). Third, the fluctuations in block energy are negatively correlated with particle density within each block inside the sample. This correlation, which also arises abruptly with the onset of a linear potential between nnn atoms, implies that excess energy fluctuations come from radially symmetric motion, i.e. a type of “breathing” mode for the atomic motions, as shown by the displacements $\delta x_{\pm}$ in Fig. 4 (B). Note that the dominant influence of this breathing mode causes nnn motion to modulate the energy of the central atom, as represented by the solid and dashed blue lines in Fig. 4 (B).

Modulation of the harmonic motion of the central atom comes primarily from the relative motion its nnn atoms. This relative motion can be separated into in-phase and out-of-phase components. These two components have well-separated frequencies, as found in the simulations. Specifically, from figure 5 in Ref. [6], the frequency of the in-phase component has a maximum of $\approx 0.5$ THz, and goes to zero with increasing block size, while the largest out-of-phase component is nearly constant at $\omega/2\pi \approx 1.5$ THz. We identify the in-phase modes with the primary heat bath while the out-of-phase modes are fast enough to be effectively decoupled from this bath, consistent with the anticorrelated energies from simulations, e.g. inset of Fig. 2 (B), and from localized modes in many measurements [9, 10, 23-33].

Now, we develop a theory for the excess fluctuations in MD simulations. Consider a time-dependent potential energy for the motion in the $x$-direction of the atom at the origin in Fig. 4. Similar contributions in the other directions are equivalent. The potential energy has a harmonic term from nn atoms, $u(x) = ax^2$, plus a time-dependent modulation, $c(t)$, from nnn atoms. Neglecting any irrelevant constant offset, the modulated potential energy is

$$u(x,t) = ax^2 + c(t). \qquad (5)$$

A linear dependence on distance ($bx$) does not appear explicitly in Eq. (5) due to the balance between the forces from opposing nnn atoms (see also Appendix A). Perfect balance requires that all nnn atoms maintain zero net displacement, $\delta x_- + \delta x_+ = 0$. However, because nnn atoms often fluctuate out-of-phase due to high-frequency vibrations that mimic zone-boundary phonons, a linear factor ($b$) from the tangential slope of the interaction between nnn atoms arises in the time-dependent modulation, $c(t) = -b \cdot (\delta x_- + \delta x_+)$. The simulations show that the out-of-phase motion of nnn atoms is dominated by a single high-frequency, $\omega$. Thus, the net displacement for the two nnn atoms on the $x$-axis in Fig. 4 (A) can be approximated by $\delta x_+ = A_+ \sin(\omega t + \alpha)$ and $\delta x_- = A_- \sin \omega t$. The amplitudes, $A_\pm$, come from the interactions between all nn atoms, which yield the plane-wave-like harmonic waves forming the heat bath at low $T$. These amplitudes can be replaced by their thermal averages, $\sqrt{\langle (A_\pm)^2 \rangle_{th}} \equiv \mathcal{A}$, in each direction. The law of equipartition of energy gives

$$\mathcal{A} = \sqrt{k_B T/(2a)}, \tag{6}$$

so that the net modulation is

$$c(t) = -b\mathcal{A}[\sin(\omega t + \alpha) + \sin \omega t]. \tag{7}$$

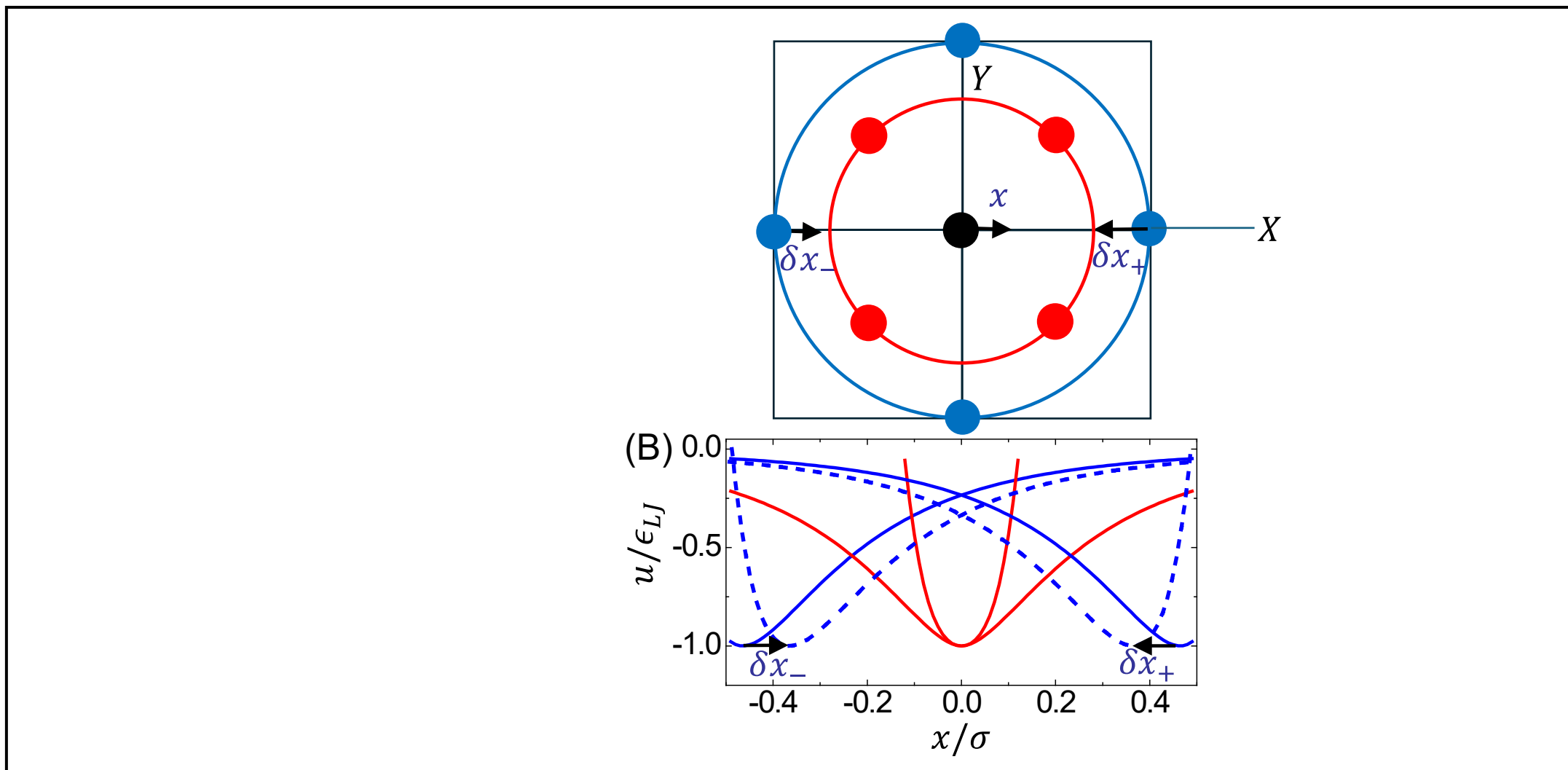

**Fig. 4.** (A) Sketch showing nine atoms in the $X$-$Y$ plane of an fcc lattice. The nn atoms (red) and nnn atoms (blue) contribute most to energy of the central atom (black) at the origin. The circles indicate the cutoff radii for nn (red) and nnn (blue) interactions. Only motions along the $X$-axis are described. $x$ and $\delta x_\pm$ are displacements of the central atom and two nnn atoms, respectively. (B) Potential energy at central atom as a function of its displacement along the $X$-axis. Solid lines are from nn atoms (red) and nnn atoms (blue). Dashed lines are from a pair of nnn atoms that are displaced by $\delta x_\pm$, showing how the energy of the central atom is modulated by the motion of its nnn atoms. All variables are dimensionless.

In contrast to the harmonic motions that are strongly correlated for nn atoms, especially for low-frequency motions, the higher-frequency breathing modes from nnn atoms are less correlated [33]. In addition, the initial phase, $\alpha$, can be treated in the random-phase approximation, combined with time averaging to obtain the equilibrium properties of the crystal.

Our theory is based on the following proposal. Equilibrium properties involve time averaging, phase averaging, and thermal averaging. Crucially, the time- and phase-averaging must be taken before thermal averaging. This ordering is natural because thermal averaging applies to static or slowly varying quantities, while the modulations are fast. (Note that for periodic functions, the long-time average is identical to the time average over one cycle.) We denote time averaging and phase averaging collectively by an overbar. After subsequent thermal averaging the mean potential energy becomes

$$\begin{aligned}\langle \bar{u} \rangle_{th} &= \frac{1}{Z}\int_{-\infty}^{\infty} dx e^{-ax^2/k_B T} \int_0^{2\pi} \frac{d\alpha}{2\pi} \int_0^{2\pi/\omega} \frac{dt}{2\pi/\omega} \{ax^2 - b\mathcal{A}[\overline{\sin(\omega t + \alpha) + \sin \omega t}]\} \\ &= \frac{1}{2} k_B T, \end{aligned} \tag{8}$$

where $Z = \int_{-\infty}^{\infty} dx e^{-ax^2/k_B T}$. Because $\overline{c(t)} = 0$, Eq. (8) yields the usual equipartition of energy, $\delta_T \langle \bar{u} \rangle_{th} / \delta T = k_B/2$. However, $\overline{c^2(t)}$ is not zero, so that the second moment of $u$ has an extra contribution:

$$\begin{aligned}\langle \overline{u^2} \rangle_{th} &= \int_{-\infty}^{\infty} dx \frac{1}{Z} e^{-ax^2/k_B T} \int_0^{2\pi} \frac{d\alpha}{2\pi} \int_0^{2\pi/\omega} \frac{dt}{2\pi/\omega} \overline{[ax^2 + c(t)]^2} \\ &= \frac{3}{4}(k_B T)^2 + b^2 \mathcal{A}^2. \end{aligned} \tag{9}$$

Then, using Eq. (6) for $\mathcal{A}$, the variance of potential energy becomes

$$(\bar{\Delta}_{th}u)^2 \equiv \langle\overline{(u - \langle\bar{u}\rangle_{th})^2}\rangle_{th} \ = \langle\overline{u^2}\rangle_{th} - \langle\bar{u}\rangle_{th}{}^2$$

$$= \frac{1}{2}(k_B T)^2 + \frac{1}{2}\frac{b^2}{a}k_B T, \tag{10}$$

which results in the modified expression for potential energy fluctuations:

$$\frac{(\bar{\Delta}_{th}u)^2}{(k_B T)^2} = \frac{1}{2} + \frac{1}{2}\frac{b^2}{a k_B T}\,. \tag{11}$$

Thus, excess fluctuations diverge from the $C \leftrightarrow (\Delta E)^2$ relation as $T \to 0$ by a factor proportional to $1/T$. Equation (8) explains why many MD simulations give good agreement with the equipartition of average energies, while Eq. (11) explains why fluctuations have an excess term, as shown in Figs. 2 (B) and 3 (B). Equation (11) also explains why excess fluctuations arise abruptly with the onset of a linear term in the potential between nnn atoms, $b \neq 0$, as found for MD simulations at low $T$ [6]. Using details from Appendix A, the linear and quadratic terms in the L-J model are found to be $a = 144\epsilon_{LJ}/\left(2^{1/3}\sigma^2\right)$ and $b = 63\epsilon_{LJ}/\left(2^{14/3}\sigma\right)$. Neglecting higher-order terms, we find the excess fluctuations to be

$$\frac{(\bar{\Delta}_{th}u)^2}{(\bar{\Delta}_{th}k)^2} - 1 = \frac{b^2}{a k_B T} = \frac{21^2}{2^{13}}\frac{\epsilon_{LJ}}{k_B T} \approx 0.053833\,\frac{\epsilon_{LJ}}{k_B T}\,. \tag{12}$$

The dashed blue line in Fig. 2 (B) shows the agreement between Eq. (12) and the excess fluctuations for individual atoms in MD simulations of the L-J model, with no adjustable parameters.

Equation (12) is for individual atoms. Most results from the MD simulations involve net fluctuations of cube-shaped blocks containing $n \gg 1$ atom. In Appendix B, we explore an analogy between nonextensive size dependence and anomalous diffusion. Specifically, a type of random walk on random walk is discussed, using a correspondence between time steps and the number of atoms in a block, $n$. Results from Appendix B yield mean-squared fluctuations in energy that vary as $(\Delta U^c)^2 = \frac{\Gamma(5/4)}{\sqrt{\pi}}(2n)^{3/4}$. When combined with Eq. (12), this gives

$$\frac{(\bar{\Delta}_{th}u)^2}{(\bar{\Delta}_{th}k)^2} - 1 \approx 0.053833\,\frac{\epsilon_{LJ} n^{3/4}/n}{k_B T}\,. \tag{13}$$

The dashed lines in Fig. 2 (B) show excellent agreement with Eq. (13), again with no adjustable parameters.

The theory yielding Eq. (13) is purely classical, as needed for MD simulations shown in Figs. 2 and 3. However, using parameters for argon, Fig. 2 (B) reveals that low-temperature deviations from the $C \leftrightarrow (\Delta E)^2$ relation begin at $T{\sim}10$ K, where quantum effects are important. In contrast, using parameters for nitromethane, figure 3(A) in Ref. [6] shows that low-temperature deviations begin near the melting temperature of $T \sim 240$ K, where classical behavior remains relevant. Nevertheless, for many measurements on most materials, quantum effects are important.

## 4. Semiclassical model for deviations from Debye's theory

We now extrapolate our classical theory to the quantum regime using a semiclassical picture. Note the qualitative similarity between experiments in Fig. 1 and simulations in Fig. 2. Figure 1 shows that measured specific heats deviate from Debye's theory as $1/T \to 0$ via $C/T^3 \propto 1/T^{\mu}$ with $\mu =$ 0.94~1.64, while Fig. 2 shows a similar deviation for excess fluctuations $\left(\Delta_{t,bl}U\right)^2/\left(\Delta_{t,bl}K\right)^2 - 1 \propto 1/T^{\mu}$ with $\mu = 1$. The simulations are purely classical, whereas low-temperature measurements involve quantum effects. We use the basic idea that semiclassical motion of nnn atoms may modulate the energies of quantized phonons in Debye's theory.

Let $\varepsilon$ be the energy of a phonon, with $\varepsilon_m$ the maximum (cut-off) energy in Debye's theory. A textbook expression [7] for the temperature-dependent part of the thermal average energy of $N$ atoms at low temperature is

$$E = \frac{9N}{\varepsilon_m^3} \int_0^{\infty} d\varepsilon\, \varepsilon^2 \left(\frac{\varepsilon}{e^{\varepsilon/k_B T} - 1}\right). \tag{14}$$

We modify Eq. (14) by treating the modulations in a "semiclassical" way as follows:

$$c_\lambda(t) = \mathcal{A}_\lambda[\sin(\omega t + \alpha) + \sin \omega t], \tag{15}$$

$$\mathcal{A}_\lambda = \kappa_\lambda (k_B T)^{\lambda/2}, \tag{16}$$

where $\kappa_\lambda$ is a $T$-independent positive constant. In the classical limit, $\lambda = 1$ and $\kappa_1 = 1/\sqrt{2a}$ in Eq. (16) yield Eq. (6) for oscillations that depend linearly on $T$ via the law of equipartition of energy. In the low-temperature limit, $\lambda = 0$ in Eq. (16) is identified with zero-point oscillations so that $\mathcal{A}_0$ is independent of $T$. Values in the range $0 < \lambda < 1$ may be identified with intermediate behavior. Although further work will be needed to fully quantize the modulations, here we examine how the semiclassical approach may qualitatively describe the measured deviations from Debye's theory.

The density of states is changed if the modulations are included, analogously to a standard treatment for Pauli paramagnetism (e.g. chapter 14 in Ref. [35]). As in our theory for classical MD simulations in Sec. 3, we assume that the modulations are decoupled from the heat bath. Such decoupled degrees of freedom are consistent with measurements of localized normal modes [23-28] and reduced correlations between nnn atoms in crystals [33].

Following the procedure used for Eq. (9), the time- and phase-averages (again denoted by an overbar) are taken before the thermal average to obtain the equilibrium behavior. Using $\hat{H}_\lambda = \varepsilon\hat{n} + c_\lambda(t)$ (where the number operator, $\hat{n}$, has eigenvalues $n = 0, 1, 2,$ ...) as the modulated single-atom Hamiltonian, Eq. (14) must be modified as

$$E_\lambda = \int_0^\infty d\varepsilon\, \overline{D(\varepsilon + c_\lambda(t))} \times \langle \overline{\hat{H}_\lambda} \rangle_{th}, \tag{17}$$

where $D(\varepsilon) = (9N/\varepsilon_m^3)\varepsilon^2$ is the density of states. Using the density matrix $\hat{\rho} = \exp(-\varepsilon\hat{n}/k_B T)/Z$ with $Z = \mathrm{tr}\exp(-\varepsilon\hat{n}/k_B T) = 1/[\exp(\varepsilon/k_B T) - 1]$, the thermal average of $\overline{\hat{H}_\lambda} = \varepsilon\hat{n}$ is $\langle \overline{\hat{H}_\lambda} \rangle_{th} = \mathrm{tr}\left(\overline{\hat{H}_\lambda}\hat{\rho}\right) = \varepsilon/\left(e^{\varepsilon/k_B T} - 1\right)$. The internal energy becomes

$$E_\lambda = \frac{9N}{\varepsilon_m^3}\int_0^\infty d\,\varepsilon\left(\varepsilon^2 + b^2\mathcal{A}_\lambda^2\right)\frac{\varepsilon}{e^{\varepsilon/k_B T} - 1}$$

$$= \frac{9N}{\varepsilon_m^3}\left[6\frac{\pi^4}{90}(k_B T)^4 + \frac{\pi^2}{6}(k_B T)^2 b^2 \mathcal{A}_\lambda^2\right]. \tag{18}$$

The dimensionless specific heat per atom becomes

$$\frac{C}{k_B} \approx \frac{9}{\varepsilon_m^3}\left[\frac{4\pi^4}{15}(k_B T)^3 + \frac{\pi^2}{6}(\lambda + 2) b^2 \kappa_\lambda^2 (k_B T)^{\lambda+1}\right]. \tag{19}$$

This result predicts an excess specific heat added to Debye's $T^3$-law, which varies as $C \propto T^{\lambda+1}$, consistent with measured excess specific heats having $0 \lesssim \lambda \lesssim 1$. Comparing this to the exponent, $\mu$, in Fig. 1, we have the relation: $\lambda + \mu = 2$. Using the classical values, $\lambda = 1$ and $\kappa_1 = 1/\sqrt{2a}$, with parameters for the argon lattice in the L-J model, $a = \left(144/2^{1/3}\right)\epsilon_{LJ}/\sigma^2$ and $b = \left(63/2^{14/3}\right)\epsilon_{LJ}/\sigma$, Eq. (19) predicts a crossover from Debye's $T^3$ behavior to excess $C \propto T^2$ at

$$T = \frac{1}{k_B}\left[\frac{5}{8\pi^2}(\lambda + 2) b^2 \kappa_1^2\right]^{1/(2-\lambda)}$$

$$= \frac{15}{16\pi^2}\left(\frac{63}{2^{14/3}}\right)^2 \frac{2^{1/3}}{144}\frac{\epsilon_{LJ}}{k_B}, \tag{20}$$

which is about 0.60 K. Measured specific heats of argon [36] down to $T \approx 0.40$ K do not exhibit any clear deviations from the $T^3$-law. However, this is not surprising because the purely classical treatment ($\lambda = 1$) is not expected to be valid at such low temperatures. Indeed, Fig. 1 shows that the measured temperature exponent of the excess specific heat of many materials is $1 \lesssim \mu \lesssim 2$, consistent with $0 \leq \lambda \leq 1$. Another factor altering the crossover is that most phonons at such low temperatures are primarily in-phase, decreasing the amplitude of the out-of-phase mode from our proposal of random atomic motions. Using interaction parameters for other crystals, such as those in Fig. 1, will clarify the connections between experiment and theory.

## 5. Conclusions

In summary, at very low temperatures, the excess specific heat measured in high-purity single crystals (Fig. 1) is similar to the excess energy fluctuations found in MD simulations of ideal crystals (Figs. 2 and 3), suggesting a common mechanism. The simulations are quantitatively characterized [dashed lines in Fig. 2 (B)] by a theory for the nonzero slope of the potential between next-nearest-neighbor atoms (Fig. 4). This theory treats independent fluctuations of these nnn atoms as a modulation on the quadratic potential between nearest-neighbor atoms. We introduce a novel method for quantifying the modulations by combining time- and phase-averaging, followed by thermal averaging. This order of averaging is justified for rapid fluctuations that do not have time to fully couple to the heat bath, as found in the simulations [Fig. 3 (C) and inset of Fig. 2 (B)] and many measurements [including Fig. 1 (A)]. Equation (11) gives a correction to the canonical-ensemble connection between fluctuations and specific heat, the $C \leftrightarrow (\Delta E)^2$ relation. We extend our theory for energy modulations in classical systems to a semiclassical model for the modulation of phonon energies in Debye's theory. The results, given in Eqs. (18) and (19), provide a qualitative explanation for the excess specific heat and excess energy fluctuations measured in glasses and crystals.

**Funding**

SA is supported by a grant from National Natural Science Foundation of China (No. 12375031).

**CRediT authorship contribution statement**

**Ralph V. Chamberlin:** Simulations, writing original draft, validation, theory development. **Sumiyoshi Abe:** Writing original draft, validation, theory development.

**Declaration of Competing Interest**

The authors declare that they have no known competing financial interests or personal relationships that could have appeared to influence the work reported in this paper.

**Acknowledgments**

RVC benefited from discussions with T. J. Newman.

*Author contributions*

RVC performed the simulations, developed the theory and wrote the original manuscript. SA developed the theory and wrote the original manuscript.

**Data availability**

Experimental data from Refs. [3-5] are available in the original publications. The simulation data are available from the corresponding author upon request.

## Appendices

### A. Linear and quadratic terms in the potential of L-J crystals

Quantitative comparison with the MD simulations in Fig. 2 can be made by evaluating the L-J model. The L-J potential energy as a function of radial distance ($r$) is $u(r) = 4\epsilon_{LJ}[(\sigma/r)^{12} - (\sigma/r)^6]$. For argon atoms, the energy scale is $\epsilon_{LJ}/k_B = 118.2$ K and the length scale is $\sigma = 0.3405$ nm [34]. The potential energy has its minimum ($u_{nn} = -\epsilon_{LJ}$) at the nn distance of $r_{nn} = 2^{1/6}\sigma$, and a value of $u_{nnn} = -15\epsilon_{LJ}/64$ at the nnn distance of $r_{nnn} = 2^{2/3}\sigma$. Expanding the potential energy to quadratic order about $r_{nn}$, the interaction between nn atoms is $u(r_{nn} + \delta r) = \epsilon_{LJ}\left[-1 + 36\,(\delta r)^2/\left(2^{1/3}\sigma^2\right)\right]$. From the geometry in Fig. 4 (A), $\delta r$ is $\pm x/\sqrt{2}$, where $x$ is the displacement from the origin of the central atom along the $X$-axis. The 4 atoms in the $Y$-$Z$ plane do not contribute, leaving 8 of the 12 nn atoms that add equally to the quadratic term. The net potential energy is

$$u_{fcc}(r_{nn} + x) = \epsilon_{LJ}\left(-12 + \frac{144}{2^{1/3}\sigma^2}x^2\right). \quad \text{(A-1)}$$

Thus, the factor for the harmonic term in the L-J potential is $a = 144\epsilon_{LJ}/\left(2^{1/3}\sigma^2\right)$. For each nnn atom at distance $r_{nnn}$, to linear order, the potential is $u(r_{nnn} + \delta r) = \epsilon_{LJ}\left[-15/64 + 21\delta r/\left(2^{14/3}\sigma\right)\right]$. In the fcc lattice there are 6 nnn atoms that can be grouped into opposing pairs displaced by $\delta x_{\pm}$, as shown in Fig. 3 (A). The effect of $\delta x_-$ from the left nnn atom is

$$u_{fcc}(r_{nnn} - \delta x_-) = \epsilon_{LJ}\left[-\frac{45}{64} - \frac{63}{2^{14/3}\sigma}\delta x_-\right], \quad \text{(A-2)}$$

and similarly the effect of $\delta x_+$ from the right nnn atom is

$$u_{fcc}(r_{nnn} - \delta x_+) = \epsilon_{LJ}\left[-\frac{45}{64} - \frac{63}{2^{14/3}\sigma}\delta x_+\right]. \quad \text{(A-3)}$$

Therefore, the net potential energy becomes

$$u_{fcc}(r_{nnn} - \delta x_-) + u_{fcc}(r_{nnn} - \delta x_+) = \epsilon_{LJ}\left[-\frac{45}{32} - \frac{63}{2^{14/3}\sigma}(\delta x_- + \delta x_+)\right]. \quad \text{(A-4)}$$

Thus, the main factor modulating the energies of the L-J model is $b = -63\epsilon_{LJ}/\left(2^{14/3}\sigma\right)$. We neglect all higher-order terms.

Due to its quadratic dependence on momentum, fluctuations in kinetic energy in the $x$-direction are $\left(\Delta_{t,bl}k\right)^2 = (k_B T)^2/2$. From Eq. (11), the excess fluctuations for the L-J model are:

$$\frac{\left(\Delta_{t,bl}u\right)^2}{\left(\Delta_{t,bl}k\right)^2} - 1 = \frac{b^2}{a k_B T}$$

$$= \frac{21^2}{2^{13}} \frac{\epsilon_{LJ}}{k_B T}$$

$$\approx 0.053833 \frac{\epsilon_{LJ}}{k_B T}. \quad \text{(A-5)}$$

Because each degree of freedom adds equally to both $k$ and $u$, Eq. (A-5) also holds in 3-dimensions. The prefactor in Eq. (A-5) is in excellent agreement with $\eta = 0.054 \pm 0.004$ from the simulations in Fig. 2 (B). This agreement implies that contributions from neighbors beyond nnn can be neglected.

**B. Anomalous diffusion of energy in blocks of atoms**

Equation (A-5) is for individual atoms. Most results from the MD simulations involve net fluctuations of cube-shaped blocks containing $n \gg 1$ atom. Denote the total kinetic and potential energies of each block by $K$ and $U$, respectively. Fluctuations of kinetic energy are extensive, $(\Delta_t K)^2 \propto n$. Equation (1) for MD simulations has a canonical-ensemble analog yielding the following $n$-dependence for excess fluctuations:

$$\frac{(\bar{\Delta}_{th}U)^2}{(\bar{\Delta}_{th}K)^2} - 1 = \eta n^{\gamma-1} \frac{\epsilon_{LJ}}{k_B T}. \quad \text{(B-1)}$$

From Eq. (1), $\gamma \approx 0.75$. The nonextensive size dependence on the right-hand side is significant over a wide range of block sizes. For the L-J model at $T = 0.1$ K, nonextensive fluctuations dominate over extensive ones if $n < 10^7$.

We propose that Eq. (B-1) can be interpreted by analogy with diffusion. If $n$ is regarded as the number of time steps in random walk, in the limit of very large systems there is normal diffusion, while the nonextensive behavior shown in Fig. 2 (B) corresponds to subdiffusion [37]. Here, we describe the nonextensive relation in terms of random walk on random walk [38,39]. For modulations of the energy of the $i$-th atom that yield the excess fluctuations, we consider the variable $U_i^c = W_i \sin \Phi_i$, where both $W_i$ and $\Phi_i$ are random, with $W_i$ made dimensionless for simplicity. We identify $(\Delta U^c)^2$ with the excess-fluctuation part, leaving the specific distributions of $W$ and $\Phi$ to be found. This form of $U^c$ is guided by Eq. (5). First, let us define $\Psi \equiv \sum_{i=1}^n \Psi_i$ ($\Psi_i = \sin \Phi_i$) and $W \equiv \sum_{i=1}^n W_i$. The joint probability of finding $\psi$ and $w$ as the realizations of $\Psi$ and $W$ at the $n$-th step is $P_n(\psi, w) = p_n(w|\psi) p_n(\psi)$. If $\Psi_i$'s are independent and identically distributed, then $\Psi$ may be assumed to obey the binomial distribution that can be approximated by the Gaussian distribution

$$p_n(\psi) = \frac{1}{2^n}\frac{n!}{\left[\frac{1}{2}(n+\psi)\right]!\left[\frac{1}{2}(n-\psi)\right]!} \approx \frac{1}{\sqrt{2\pi n}}\,\exp\left(-\frac{\psi^2}{2n}\right). \tag{B-2}$$

The joint distribution and Eq. (B-2) are integrated over a range of values to obtain the net mean-squared deviations of $U^c$. Assuming large enough block sizes (but still $n < 10^7$), the range of integration can be extended to $(-\infty, \infty)$. Since the distributions are even functions, the average of $U^c$ vanishes. Then, if the conditional probability distribution is taken to be

$$p(w|\psi) = \frac{|\psi|^{1/4}}{\sqrt{2\pi}}\exp\left(-\frac{\sqrt{|\psi|}w^2}{2}\right), \tag{B-3}$$

which is independent of $n$, mean-squared fluctuations become

$$(\Delta U^c)^2 = \int_{-\infty}^{\infty} d\psi \int_{-\infty}^{\infty} dw\, \psi^2 w^2\, p_n(\psi) p_n(w|\psi)$$

$$= \frac{\Gamma(5/4)}{\sqrt{\pi}}(2n)^{3/4}, \tag{B-4}$$

Although the dashed lines in Fig. 2 (B) show good agreement with the size dependence of Eq. (B-4), the conditional probability in Eq. (B-3) should be further validated by future work.